\documentclass[showpacs,pra]{revtex4}
\usepackage{amsmath}
\usepackage{graphicx}
\begin{document}
\title{Vortex lattice solutions to the Gross-Pitaevskii equation with spin-orbit coupling in optical lattices}
\author{Hidetsugu Sakaguchi and Ben Li}
\affiliation{Department of Applied Science for Electronics and Materials,
Interdisciplinary Graduate School of Engineering Sciences, Kyushu
University, Kasuga, Fukuoka 816-8580, Japan}
\begin{abstract}
Effective spin-orbit coupling can be created in cold atom systems using atom-light interaction. We study the BECs in an optical lattice using the Gross-Pitaevskii equation with spin-orbit coupling. 
Bloch states for the linear equation are numerically obtained, and compared with stationary solutions to the Gross-Pitaevskii equation with nonlinear terms. 
Various vortex lattice states are found when the spin-orbit coupling is strong. \end{abstract}
\maketitle
Recently, Bose-Einstein condensates (BECs) with effective spin-orbit coupling were created in cold atom systems using atom-light interaction~\cite{rf:1}. The spin-orbit-coupled BECs are actively studied theoretically~\cite{rf:2}. Wang et al. found that the mean-field ground state has two different phases: plane-wave and stripe phases depending on the nonlinear interactions~\cite{rf:3}. Half vortex states were found in a spin-orbit coupled BECs confined in a harmonic potential~\cite{rf:4,rf:5}. Exotic spin textures were predicted in Bose-Hubbard models corresponding to spin-orbit coupled BECs in the Mott-insulator phase~\cite{rf:6,rf:7}. 

The GRoss-Pitaevskii (GP) equation is a mean-field approximation for the BECs with the spin-orbit coupling. There are some studies for the GP equation with spin-orbit coupling in optical lattices~\cite{rf:8,rf:9}. In this paper, we study vortex lattice solutions to the GP equation in a square type optical lattice. 
The model equation is expressed as
\begin{eqnarray}
i\frac{\partial \psi_{+}}{\partial t}&=&-\frac{1}{2}\nabla^2\psi_{+}+(g|\psi_{+}|^2+\gamma|\psi_{-}|^2)\psi_{+}-\epsilon\{\cos(2\pi x)+\cos(2\pi y)\}\psi_{+}+\lambda\left (\frac{\partial \psi_{-}}{\partial x}-i\frac{\partial \psi_{-}}{\partial y}\right ),\nonumber\\
i\frac{\partial \psi_{-}}{\partial t}&=&-\frac{1}{2}\nabla^2\psi_{-}+(g|\psi_{-}|^2+\gamma|\psi_{+}|^2)\psi_{-}-\epsilon\{\cos(2\pi x)+\cos(2\pi y)\}\psi_{-}+\lambda\left (-\frac{\partial \psi_{+}}{\partial x}-i\frac{\partial \psi_{+}}{\partial y}\right ),
\end{eqnarray}
where $\boldsymbol{\psi}=(\psi_{+},\psi_{-})$ denotes the wave function of the spinor BECs, $\epsilon$ is the strength of the optical lattice, $g$ and $\gamma$ express the strengths of interactions respectively between the same and the different kinds of atoms, and $\lambda$ denotes the strength of the Rashba spin-orbit coupling. We have assumed that the wavelength of the optical lattice is 1.

If $g$ and $\gamma$ are zero, Eq.~(1) becomes linear equations with spatially-periodic potential. The Bloch states are stationary solutions to the linear equations, which are expressed as
\begin{equation}
\psi_{+}(x,y,t)=\phi_{+}(x,y)\exp(ik_xx+ik_yy-i\mu t),\;\psi_{-}(x,y,t)=\phi_{-}(x,y)\exp(ik_xx+ik_yy-i\mu t),
\end{equation}
where $\phi_{+}$ and $\phi_{-}$ are periodic functions of wavelength $1$.
Therefore, $\phi_{+}$ and $\phi_{-}$ satisfy
\begin{eqnarray}
\mu\phi_{+}&=&-\frac{1}{2}\nabla^2\phi_{+}+\frac{1}{2}(k_x^2+k_y^2)\phi_{+}-ik_x\frac{\partial\phi_{+}}{\partial x}-ik_y\frac{\partial \phi_{+}}{\partial y}-\epsilon\{\cos(2\pi x)+\cos(2\pi y)\}\phi_{+}\nonumber\\
& &+\lambda\left (\frac{\partial \phi_{-}}{\partial x}-i\frac{\partial \phi_{-}}{\partial y}+ik_x\phi_{-}+k_y\phi_{-}\right ),\nonumber\\
\mu\phi_{-}&=&-\frac{1}{2}\nabla^2\phi_{-}+\frac{1}{2}(k_x^2+k_y^2)\phi_{-}-ik_x\frac{\partial\phi_{-}}{\partial x}-ik_y\frac{\partial \phi_{-}}{\partial y}-\epsilon\{\cos(2\pi x)+\cos(2\pi y)\}\phi_{-}\nonumber\\
& &+\lambda\left (-\frac{\partial \phi_{+}}{\partial x}-i\frac{\partial \phi_{+}}{\partial y}-ik_x\phi_{+}+k_y\phi_{+}\right ).
\end{eqnarray}
The eigenvalue $\mu$ and the eigen function $\phi_{+}$ and $\phi_{-}$ can be numerically obtained from the stationary solution of the linear equation~\cite{rf:10}:
\begin{eqnarray}
\frac{\partial \phi_{+}}{\partial t}&=&\frac{1}{2}\nabla^2\phi_{+}-\frac{1}{2}(k_x^2+k_y^2)\phi_{+}+ik_x\frac{\partial\phi_{+}}{\partial x}+ik_y\frac{\partial \phi_{+}}{\partial y}+\epsilon\{\cos(2\pi x)+\cos(2\pi y)\}\phi_{+}\nonumber\\
& &-\lambda\left (\frac{\partial \phi_{-}}{\partial x}-i\frac{\partial \phi_{-}}{\partial y}+ik_x\phi_{-}+k_y\phi_{-}\right )+\mu\phi_{+},\nonumber\\
\frac{\partial\phi_{-}}{\partial t}&=&\frac{1}{2}\nabla^2\phi_{-}-\frac{1}{2}(k_x^2+k_y^2)\phi_{-}+ik_x\frac{\partial\phi_{-}}{\partial x}+ik_y\frac{\partial \phi_{-}}{\partial y}+\epsilon\{\cos(2\pi x)+\cos(2\pi y)\}\phi_{-}\nonumber\\
& &-\lambda\left (-\frac{\partial \phi_{+}}{\partial x}-i\frac{\partial \phi_{+}}{\partial y}-ik_x\phi_1+k_y\phi_{+}\right )+\mu\phi_{-},\nonumber\\
\frac{d\mu}{dt}&=&\alpha(N_0-N),
\end{eqnarray}
where $\alpha>0$ is a parameter and fixed to be 5 in our numerical simulation. $N=\int_0^1\int_0^1(|\phi_{+}|^2+|\phi_{-}|^2)dxdy$ is the total norm, and  $N_0$ is fixed to be 1 by the normalization condition. The time evolution of the dissipative equation (4) leads to a stationary state and the total norm $N$ approaches $N_0=1$. The eigenvalue $\mu$ in Eq.~(3) is obtained as $\mu$ in Eq.~(4) at the stationary state. In this numerical method, the ground state for fixed values of $k_x$ and $k_y$ is obtained at the stationary state, starting from most initial conditions, because the total energy decreases in the time evolution of Eq.~(4). Excited states are obtained by removing the ground state by the method of orthogonalization. 
 Figure 1(a) shows $\mu(k_x)$ as a function of $k_x$ for $k_y=0,\epsilon=5$, and $\lambda=\pi/2$.  $\mu(k_x)$ is a periodic function of $k_x$ with period $2\pi$. There are peaks near $k_x=0, \pi$ and $2\pi$ and minima at $k_x\sim \pi/2$ and $3\pi/2$. 
The peak point at $k_x=\pi$ is a cusp point, where two $\mu(k_x)$ curves corresponding to the ground state and the excited state cross, although the branch of the excited state is not shown. For $\lambda=0$, $\mu(k_x)$ increases monotonously as $k_x$ increases from 0 and reaches the maximum at the edge of the Brillouin zone at $k_x=\pi$. If there is no optical lattice, i.e., $\epsilon=0$, $\mu(k)$ takes a minimum at $k=\lambda$ where $k=\sqrt{k_x^2+k_y^2}$~\cite{rf:2,rf:3}. The minimum point for $\lambda=\pi/2$ locates near $k_x=\lambda$, and the peak corresponds to the edge of the Brillouin zone. Figure 1(b) shows $\mu(k_x)$ at $\lambda=(3/2)\pi$.  There are a large peak at $k_x=0$ and $2\pi$ and a small peak at $k_x=\pi$ and minima at $k_x=2$ and $k_x=4.3$.  Figure 1(c) shows $\mu(k)$ as a function of $k_x$ for $k_x=k_y,\lambda=\pi,\epsilon=5$.  There is a large peak at $k_x=0$ and $2\pi$, a small peak at $k_x=\pi$, and minima at $k_x\sim 2.5$ and 3.8. 
The wavenumber $k_x\sim 2.5$ is close to $\lambda/\sqrt{2}\sim 2.22$ by the simplest approximation $k=\lambda$  but slightly deviated. The approximation $k_{{\rm min}}=\lambda$ for the minimum point of $\mu(k)$ becomes worse for large $\lambda$. The small peaks correspond to the edge of the Brillouin zone. 
\begin{figure}[tbp]
\begin{center}
\includegraphics[height=4.cm]{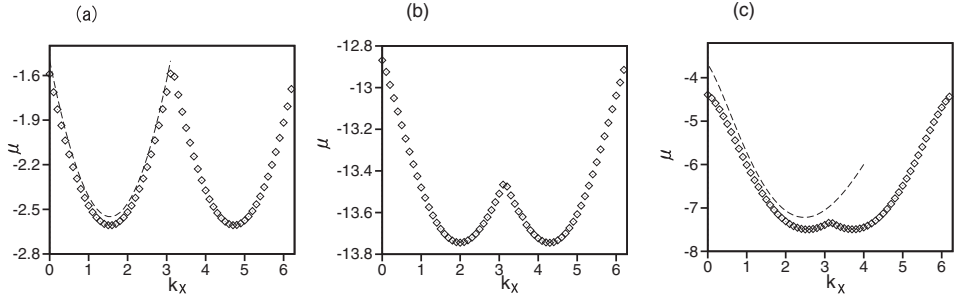}
\end{center}
\caption{Eigenvalue $\mu$ vs. $k_x$ for (a) $\lambda=\pi/2, k_y=0$, (b) $\lambda=3\pi/2, k_y=0$, and (c) $\lambda=\pi, k_y=k_x$. Dashed curve in Fig.1(a) is plotted using Eq.~(8) and the dashed curve in Fig.~1(c) is obtained using Eq.~(6).}
\label{f1}
\end{figure}

Figure 2(a) shows $|\phi_{+}(x,y)|$ and $|\phi_{-}(x,y)|$ as a function of $y$ in the section $x=0$ at $k_x=3\pi/2$ for $\lambda=3\pi/2$. The modulus $|\phi_{+}|$ and $|\phi_{-}|$ take maximum at different positions. The minimum value is almost zero, which implies the existence of vortices. Figure 2(b) shows a contour plot of $|\phi_{+}|$ for the same parameter. 
The locations of vortices for $\phi_{+}$ can be calculated from the phase distribution $\theta_{+}(x,y)=\sin^{-1}({\rm Im}\, \phi_{+}(x,y)/|\phi_{+}(x,y)|)$. There exist a vortex at a point, if the integral of the phase grandient along an anticlockwise path encircling the point is a nontrivial multiple of $2\pi$. We have counted the path integral by discretizing the $(x,y)$ space with $\Delta x=1/64$. Figure 2(c) shows positions of vortices of vorticity $\pm 1$ with square and $\times$ marks. 
The vortex cores locate near $(0,-0.28)$ and $(0,-0.48)$ for $\phi_{+}$. In generic cases, there is a vortex of vorticity 1 or -1 
 at a position satisfying $|\phi_+|=0$, where a line of ${\rm Re}\, \phi_{+}=0$ intersects with a line of ${\rm Im}\,\phi_{+}=0$~\cite{rf:11}. 
We do not show explicitly the positions of vortices later in Fig.~3 and Fig.~4, however, we have checked the existence of vortices of vorticity 1 or -1 at positions satisfying $|\phi_{\pm}|=0$ by calculating the phase distribution. 
Figure 2(d) shows the minimum value of $|\phi_{+}|$ as a function of $\lambda$ for $k_x=\lambda,k_y=0$ at $\epsilon=5$. The minimum value becomes zero and a vortex-antivortex pair appears for $\lambda>\lambda_c\sim 4.2$. 

Because $\phi_{\pm}$ are periodic functions with wavelength 1, $\phi_{\pm}$ can be expressed as the simplest approximation: 
\begin{eqnarray}
\phi_{+}&=&C_{0+}+C_{1+}e^{2\pi i x}+C_{2+}e^{-2\pi ix}+C_{3+}e^{2\pi iy}+C_{4+}e^{-2\pi iy},\nonumber\\
\phi_{-}&=&C_{0-}+C_{1-}e^{2\pi i x}+C_{2-}e^{-2\pi ix}+C_{3-}e^{2\pi iy}+C_{4-}e^{-2\pi iy}.
\end{eqnarray}
\begin{figure}[tbp]
\begin{center}
\includegraphics[height=4.cm]{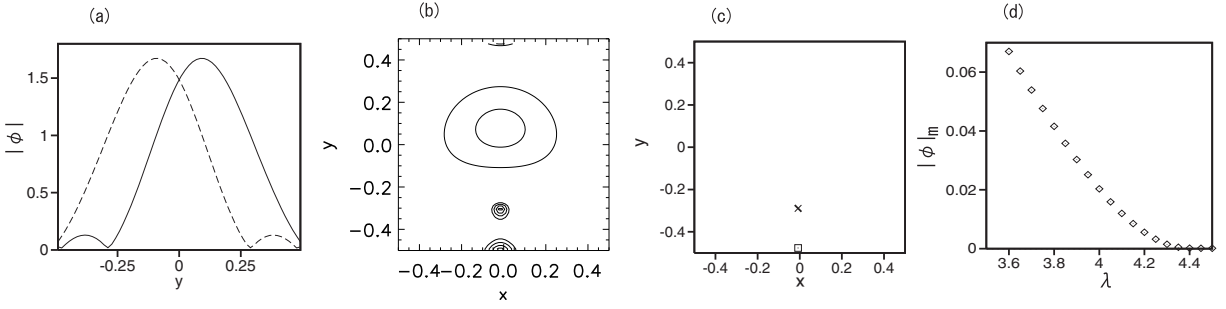}
\end{center}
\caption{(a) $|\phi_{+}|$ (solid curve) and $|\phi_{-}|$ (dashed curve) along the line $x=0$ for $\lambda=3\pi/2$, and $k_x=3\pi/2$. (b) Contour plot of $|\phi_{+}|$. (c) Square shows a vortex with vorticity 1 and $\times$ shows a vortex with vorticity -1. (d) Minimum values of $|\phi_{+}|$ as a function of $\lambda$ for $k_x=\lambda$ and $k_y=0$.}
\label{f2}
\end{figure}
Substitution of this ansatz into Eq.~(3) yields 
\begin{eqnarray}
\mu C_{0\pm}&=&(k_x^2+k_y^2)C_{0\pm}/2-(\epsilon/2)(C_{1\pm}+C_{2\pm}+C_{3\pm}+C_{4\pm})+\lambda(\pm ik_x+k_y)C_{0\mp},\nonumber\\
\mu C_{1\pm}&=&\{(k_x+2\pi)^2+k_y^2\}C_{1\pm}/2-(\epsilon/2)C_{0\pm}+\lambda\{\pm i(k_x+2\pi)+k_y\}C_{1\mp},\nonumber\\
\mu C_{2\pm}&=&\{(k_x-2\pi)^2+k_y^2\}C_{2\pm}/2-(\epsilon/2)C_{0\pm}+\lambda\{\pm i(k_x+2\pi)+k_y\}C_{2\mp},\nonumber\\
\mu C_{3\pm}&=&\{k_x^2+(k_y+2\pi)^2\}C_{3\pm}/2-(\epsilon/2)C_{0\pm}+\lambda\{\pm ik_x+(k_y+2\pi)\}C_{3\mp}\nonumber\\
\mu C_{4\pm}&=&\{k_x^2+(k_y-2\pi)^2\}C_{4\pm}/2-(\epsilon/2)C_{0\pm}+\lambda\{\pm ik_x+(k_y-2\pi)\}C_{4\mp}.
\end{eqnarray}
For $k_y=0$, $C_{0-}=iC_{0+},C_{1-}=iC_{1+},C_{2-}=iC_{2+}$ are satisfied, and 
then
\begin{eqnarray}
C_{1+}&=&\frac{-(\epsilon/2)C_{0+}}{\mu-(k_x+2\pi)^2/2+\lambda(k_x+2\pi)},\nonumber\\
C_{2+}&=&\frac{-(\epsilon/2)C_{0+}}{\mu-(k_x-2\pi)^2/2+\lambda(k_x-2\pi)},\nonumber\\
C_{3+}&=&\frac{[-(\epsilon/2)\{\mu-(k_x^2+4\pi^2)/2)\}+(\epsilon/2)\lambda (k_x-2\pi i)]C_{0+}}{\{\mu-(k_x^2+4\pi^2)/2\}^2-\lambda^2(k_x^2+4\pi^2)},\nonumber\\
C_{4+}&=&\frac{[-(\epsilon/2)\{\mu-(k_x^2+4\pi^2)/2)\}+(\epsilon/2)\lambda (k_x+2\pi i)]C_{0+}}{\{\mu-(k_x^2+4\pi^2)/2\}^2-\lambda^2(k_x^2+4\pi^2)},
\end{eqnarray}
where $\mu$ is given by a solution of the equation
\begin{eqnarray}
\mu&=&\frac{k_x^2}{2}-\lambda k_x+\frac{\epsilon^2/4}{\mu-(k_x+2\pi)^2/2+\lambda(k_x+2\pi)}+\frac{\epsilon^2/4}{\mu-(k_x-2\pi)^2/2+\lambda(k_x-2\pi)}\nonumber\\& &+\frac{\epsilon^2}{4}\frac{2\mu-(k_x^2+4\pi^2)-2\lambda k_x}{\{\mu-(k_x^2+4\pi^2)/2\}^2-\lambda^2(k_x^2+4\pi^2)}.
\end{eqnarray}
Furthermore, $C_{3-}=iC_{3+}^{*},C_{4-}=iC_{4+}^*$ are satisfied.   Here, $^{*}$ denotes the complex conjugate. 
The dashed curve in Fig.~1(a) denotes $\mu(k_x)$ by Eq.~(8) at $\lambda=\pi/2$. The approximation is good for $ \lambda=\pi/2$ but is not so good for large $\lambda$, because the higher harmonics is necessary for the expansion in Eq.~(5).  We can assume that $C_{0+},C_{1+}$ and $C_{2+}$ are real numbers and $C_{4+}=C_{3+}^{*}={\rm Re}C_{3+}-i{\rm Im}C_{3+}$. Then, $\phi_{+}$ and $\phi_{-}$ are expressed as
\begin{eqnarray}
\phi_{+}&=&C_{0+}+(C_{1+}+C_{2+})\cos(2\pi x)+i(C_{1+}-C_{2+})\sin(2\pi x)+2{\rm Re}C_{3+}\cos(2\pi y)-2{\rm Im}C_{3+}\sin(2\pi y),\nonumber\\  
\phi_{-}&=&iC_{0+}+i(C_{1+}+C_{2+})\cos(2\pi x)-(C_{1+}-C_{2+})\sin(2\pi x)+2i{\rm Im}C_{3+}\sin(2\pi y)+2i{\rm Re}C_{3+}\cos(2\pi y).
\end{eqnarray} 
${\rm Im}\,\phi_{+}=0$ and ${\rm Re}\,\phi_{-}=0$ are satisfied on the line $x=0$. 
When $\lambda$ is small, the minimum values of ${\rm Re}\,\phi_{+}$ and ${\rm Im}\,\phi_{-}$ are positive and there are no vortices. When $\lambda$ is increased the minimum values decrease and reach 0, and then a vortex pair is created.  
A vortex core of $\phi_{+}$ is located at a point on the line $x=0$ where ${\rm Re}\,\phi_{+}(0,y)=0$ is satisfied, and similarly a vortex core of $\phi_{-}$ is located at a point on the line $x=0$ where ${\rm Im}\,\phi_{-}(0,y)=0$ is satisfied. 

\begin{figure}[tbp]
\begin{center}
\includegraphics[height=4.5cm]{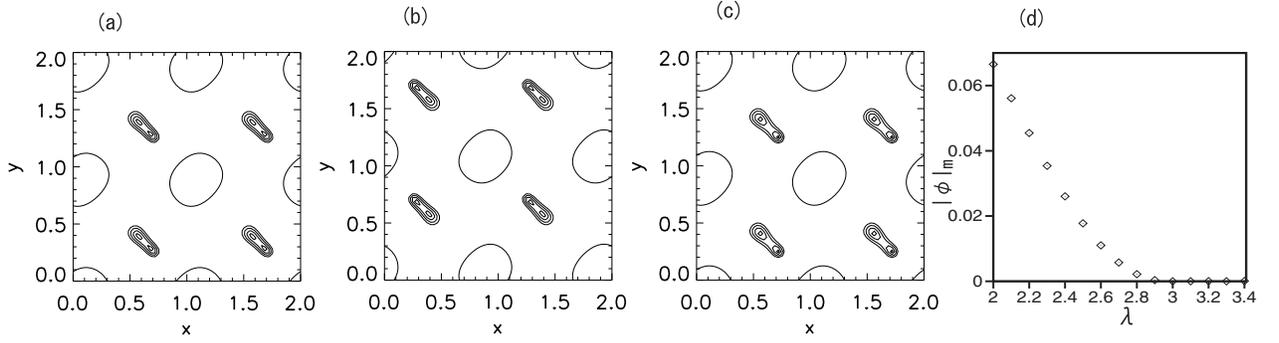}
\end{center}
\caption{(a) Contour plot of $|\psi_{+}|$ for $\lambda=\pi,g=1,\gamma=0.5$ and $L=8$. $k_x=k_y$ are evaluated as $3\pi/4$. (b) Contour plot of $|\psi_{-}|$. (c) Contour plot of $|\phi_{+}|$ to the linear equation Eq.~(3) for $\lambda=\pi, k_x=k_y=3\pi/4$. (d) Minimum values of $|\phi_{+}|$ as a function of $\lambda$ for $k_x=k_y=\lambda/\sqrt{2}$.}
\label{f3}
\end{figure}
\begin{figure}[tbp]
\begin{center}
\includegraphics[height=4.5cm]{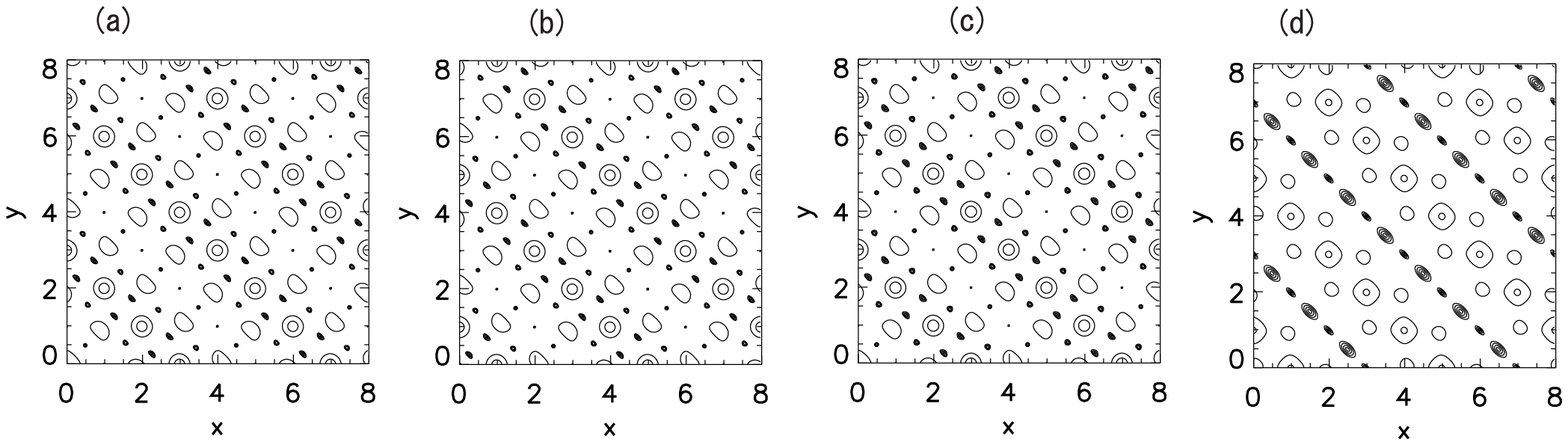}
\end{center}
\caption{(a) Contour plot of $|\psi_{+}|$ for $\lambda=\pi,g=1,\gamma=2$ and $L=8$. $k_x=k_y$ are evaluated as $3\pi/4$. (b) Contour plot of $|\psi_{-}|$. (c) Contour plot of the superposition of $|(\phi_{++}+\phi_{+-})/\sqrt{2}|$ to the linear equation Eq.~(3) for $\lambda=\pi$ and $k_x=k_y=\pm 3\pi/4$. (d) Contour plot of the superposition of $|(\phi_{++}+\phi_{+-})/\sqrt{2}|$ to the linear equation Eq.~(3) for $\lambda=1$ and $k_x=k_y=\pm \pi/4$}
\label{f4}
\end{figure}
Even for $g$ and $\gamma$ is not zero, the Bloch state is a good approximation for the stationary state for $\gamma<g$. We have performed numerical simulation of Eq.~(1) by the imaginary time evolution method similar to Eq.~(4) and found stationary solutions. The system size is $L_x\times L_y=L\times L$ and the total norm $N=\int_0^{L}\int_0^{L}(|\psi_{+}|^2+|\psi_{-}|^2)dxdy$ is set to be $L^2$ in this paper. Periodic boundary conditions are imposed. 
The potential is shifted as $U=-\epsilon[\cos\{2\pi(x-1/2)\}+\cos\{2\pi(y-1/2)\}]$ by $(1/2,1/2)$ to confine the wave pattern in the range of $[0,L]\times[0,L]$. 

Figure 3(a) and (b) show contour plots of $|\psi_{+}|$ and $|\psi_{-}|$ at $g=1,\gamma=0.5, L=8, \lambda=\pi$, and $\epsilon=5$. The contour plot is drawn in the region $[0,2]\times [0,2]$, and the contour lines are drawn for $|\psi_{\pm}|=0.025,0.05,0.075,0.1,1$ and 1.5. Vortex pairs exist in each cell of size 1 for this parameter, and a vortex lattice is constructed as a whole. Vortex lattices were experimentally found first in rotating BECs~\cite{rf:12} and recently in BECs under synthetic magnetic fields by atom-light interaction~\cite{rf:13}. In our model equation, vortices are spontaneously created by the spin-orbit coupling. 
The wavevector $(k_x,k_y)$ is evaluated at $(3\pi/4,3\pi/4)$. Positions of vortex cores for $\psi_{+}$ and $\psi_{-}$ are mutually deviated.  Figure 3(c) shows a contour plot of $|\phi_{+}|$ for the linear equation corresponding to  $g=0,\gamma=0$ for $k_x=k_y=3\pi/4$ at $\lambda=\pi$ and $\epsilon=5$. 
The eigenvalue $\mu$ takes a minimum at $(k_x,k_y)=(3\pi/4,3\pi/4)$ in the finite size system of $L=8$, where $k_x$ ($k_y$) takes a discrete value $2\pi n_x/L$ ($2\pi n_y/L$) with integer $n_x$ ($n_y$).  
  The contour plot is almost the same as Fig.~3(a). It means that the Bloch wave is a good approximation for the solution to the GP equation. Figure 3(d) shows the minimum values of $|\phi_{+}|$ for the linear equation as a function of $\lambda$ for $k_x=k_y=\lambda/\sqrt{2}$ at $\epsilon=5$. The minimum value becomes zero and vortices appear for $\lambda>2.9$. It is related to the existence of vortices at $\lambda=\pi$. 

Stripe wave states are expected to appear for $\gamma>g$. The superposition of Bloch waves of $(k_x,k_y)$ and $(-k_x,-k_y)$ is a simple    approximation for $\gamma>g$. Figure 4(a) and (b) show contour plots of $|\psi_{+}|$ and $|\psi_{-}|$ at $g=1,\gamma=2, L=8$, and $\lambda=\pi$. The wavevector is evaluated as $(k_x,k_y)=(3\pi/4,3\pi/4)$ in this case, too. 
The contour lines are drawn for $|\psi_{\pm}|=0.025,0.05,0.075,0.1,1$ and 1.5. 
Vortex cores exist in dark pointed regions. The vortex lattice structure is rather complicated. The circular contour lines correspond to peak regions of $|\psi_{\pm}|$.  The peak regions stand in a line in the direction of angle $-\pi/4$
 and the peak lines for $\psi_{+}$ and $\psi_{-}$ alternates in the diagonal direction of angle $\pi/4$.  Figure 4(c) shows a contour plot of a linear combination $|(\phi_{++}+\phi_{+-})/\sqrt{2}|$ of two Bloch waves $\phi_{++}$ and $\phi_{+-}$ with $(k_x,k_y)=(3\pi/4,3\pi/4)$ and $(-3\pi/4,-3\pi/4)$ for the $+$ component at $\lambda=\pi$.  The superposition of the Bloch waves is a good approximation for the stationary solution to the GP equation. The superposition of two Bloch waves with opposite wavevectors generates a standing wave. For plane waves, the amplitude becomes zero at the nodal lines. The nodal lines are perturbed by the optical lattice and vortices are generated. A vortex lattice structure therefore appears even for small $\lambda$ in case of $\gamma>g$. Figure 4(d) shows a vortex lattice pattern with $k_x=k_y=\pi/4$ at $\lambda=1$ and $\epsilon=5$. 
For large $\lambda$, a vortex pair is created in a single Bloch wave and the superposition of the two Bloch waves make the vortex lattice structure more complicated as shown in Fig.~4(c). 

The complicated patterns might be simplified, if a spin representation is used, which was discussed in the Bose-Hubbard model~\cite{rf:6,rf:7}. The whole system is divided into cell regions of $[i-1,i]\times[j-1,j]$. 
The spin variables $s_x(i,j),s_y(i,j)$ and $s_z(i,j)$ are  defined for each cell labeled by $(i,j)$ as
\begin{eqnarray}
s_x(i,j)&=&\int_{i-1}^{i}\int_{j-1}^{j} \boldsymbol\psi^{\dag}\sigma_x\boldsymbol\psi dxdy=\int_{i-1}^{i}\int_{j-1}^{j}(\psi_{+}^*\psi_{-}+\psi_{-}^*\psi_{+})dxdy,\nonumber\\
s_y(i,j)&=&\int_{i-1}^{i}\int_{j-1}^{j} \boldsymbol\psi^{\dag}\sigma_y\boldsymbol\psi dxdy=\int_{i-1}^{i}\int_{j-1}^{j}(-i\psi_{+}^*\psi_{-}+i\psi_{-}^*\psi_{+})dxdy,\nonumber\\
s_z(i,j)&=&\int_{i-1}^{i}\int_{j-1}^{j} \boldsymbol\psi^{\dag}\sigma_z\boldsymbol\psi dxdy=\int_{i-1}^i\int_{j-1}^j(|\psi_{+}|^2-|\psi_{-}|^2)dxdy,
\end{eqnarray}
where $\sigma_x,\sigma_y$ and $\sigma_z$ are the Pauli matrix, and $^{\dag}$ denotes the complex conjugate transpose. 
\begin{figure}[tbp]
\begin{center}
\includegraphics[height=4.5cm]{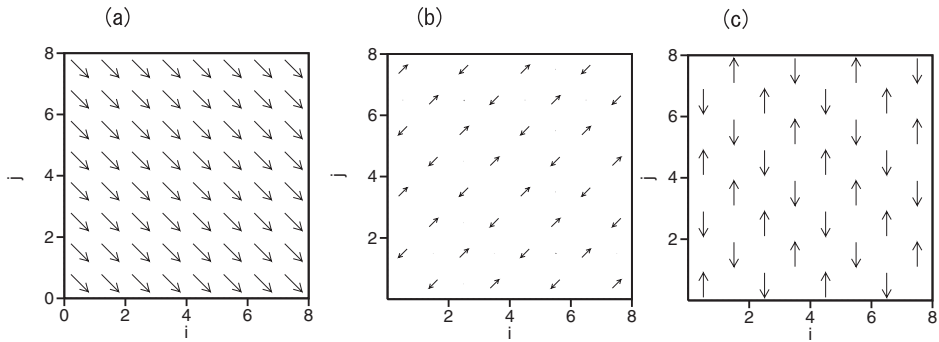}
\end{center}
\caption{(a) Spin configuration of $(s_x(i,j),s_y(i,j))$ at $\lambda=\pi,g=1,\gamma=0.5$ and $L=8$. (b) Spin configuration of $(s_x(i,j),s_y(i,j))$ at $\lambda=\pi,g=1,\gamma=2$ and $L=8$. (c) Spin configuration of $s_z(i,j)$ at $\lambda=\pi,g=1,\gamma=2$ and $L=8$.}  
\label{f5}
\end{figure}
Figure 5(a) shows $(s_x,s_y)$ corresponding to the pattern in Figs.~3(a) and (b) for $g=1,\gamma=0.5,\lambda=\pi$ and $\epsilon=5$. 
The vector $(s_x(i,j),s_y(i,j))$ is expressed as an arrow on each lattice point  at $(i-1/2,j-1/2)$. 
The pattern is interpreted as a ferromagnetic state in the $(x,y)$ plane in this spin representation. The spin $s_z$ is zero for this pattern. 
Figures~5(b) and  (c) show spin configurations respectively for $(s_x,s_y)$ and  $s_z$ for the pattern at $g=1$ and $\gamma=2$ shown in Figs.~4(a) and (b). The spin configuration is also rather complicated. 
The wavelength of the spin configuration is 4 both in the $i$ and $j$ directions. An anti-ferromagnetic order is seen in the diagonal direction of angle $\pi/4$ and a ferromagnetic order appears in its orthogonal direction of angle $-\pi/4$  both for the $(s_x,s_y)$ and $s_z$ patterns. The $(s_x,s_y)$ component appears at the sites where the $s_z$ component vanishes, and the $s_z$ component appears at the sites where the $(s_x,s_y)$ component vanishes.  

To summarize, we have studied the Gross-Pitaevskii equation with spin-orbit coupling in an optical lattice. We have found that a vortex lattice structure appears for large $\lambda$ in case of $\gamma<g$. A vortex lattice structure appears even for small $\lambda$ in case of $\gamma>g$, because the nodal lines in the stripe wave pattern are perturbed by the optical lattice. We have found a complicated spin configuration in a case of $\gamma>g$. The complicated patterns can be qualitatively understood by the corresponding Bloch waves. The Bloch waves are further approximated by a Fourier series expansion with five modes to understand the formation of the vortices. We have obtained various spin configurations by changing the parameter $\lambda$. The detailed phase diagrams by changing various parameters are under study. 

\end{document}